\documentclass[12pt]{iopart}
\usepackage{graphicx}
\usepackage{amssymb}
\usepackage{iopams}
\usepackage{setstack}

\begin{document}

\title{Phase Estimation from Atom Position Measurements}
\author{J Chwede\'nczuk$^{1}$, F Piazza$^{2}$ and A Smerzi$^{2}$}
\address{
  $^1$Faculty of Physics, University of Warsaw, Poland\\
  $^2$INO-CNR, BEC Center, Via Sommarive 14, 38123 Povo, Trento, Italy}

\begin{abstract} 
  We study the measurement of the position of atoms as a means to estimate the relative phase between two Bose-Einstein condensates. First, we consider $N$ atoms released 
  from a double-well trap, forming an interference pattern, and show that a simple least-squares fit to the density gives a shot-noise limited sensitivity. The shot-noise limit can instead be overcome by using correlation functions of order $\sqrt{N}$ or larger. The measurement of the $N\mathrm{th}$-order correlation function allows to estimate the relative phase at the Heisenberg limit. 
 Phase estimation through the measurement of the center-of-mass of the interference
  pattern can also provide sub-shot-noise sensitivity. 
  Finally, we study the effect of the overlap between the two clouds on the phase estimation, when Mach-Zehnder interferometry is performed in a double-well. 
  We find that a non-zero overlap between the clouds dramatically reduces the phase sensitivity. 
\end{abstract}
\maketitle

\section{Introduction}
Interferometry aims at the estimation of the relative phase between two 
wave-packets. In a standard optical interferometer, like the well known Mach-Zehnder setup \cite{mach}, these two wave-packets correspond to the light traveling 
inside the two arms of the device, and the relative phase $\theta$ is acquired, for instance, as a result of different optical path length. 
After the phase is accumulated, the two wave-packets are recombined through a beam-splitter, and the signal at the two output ports depends on $\theta$. 
The phase can be estimated by measuring the difference in intensities between these ports.
Apart from photons, atoms can be employed for interferometric purposes as well \cite{cronin}. The atoms present some interesting advantages with respect to light, 
especially due to the non-zero mass. In particular, the creation of atomic Bose-Einstein condensates (BECs) opened a new chapter in the field of interferometry. 
The BEC, which behaves like a macroscopic matter-wave, constitutes a coherent
and well-controllable source of particles.
This makes the BEC a promising system to measure the electromagnetic \cite{cornell,ketterle_casimir,vuletic_casimir} or
gravitational \cite{hinds,fattori,kasevich} forces. Moreover, the 
inter-atomic interactions in the BEC are a source of nonlinearity, which can be used to create non-classical states 
\cite{esteve,  GrossNature2010, treutlein,MaussangArxiv2010} useful to overcome the limit imposed by the classical physics on measurement precision \cite{giovanetti,pezze}.

A BEC interferometer can be implemented using a double-well trap \cite{shin,schumm,AlbiezPRL2005,LevyNature2007,toronto,pezze2005,lee,huang,grond1}, where the two wave-packets are localized about the two minima of the external potential. 
In such configuration, a relative phase $\theta$ can be accumulated by letting the system evolve in time in presence of an energy difference between the two potential minima, 
and in absence of coupling between the two wells. After this stage, one can, for example, recombine the wave-packets by implementing a beam splitter 
(thereby realizing a Mach-Zehnder interferometer). 
This will imply a further dynamical evolution during which atoms oscillate between the wells for a time which must be precisely under control, and over which interactions are negligible. 

\begin{figure}
  \includegraphics[scale=.48,clip]{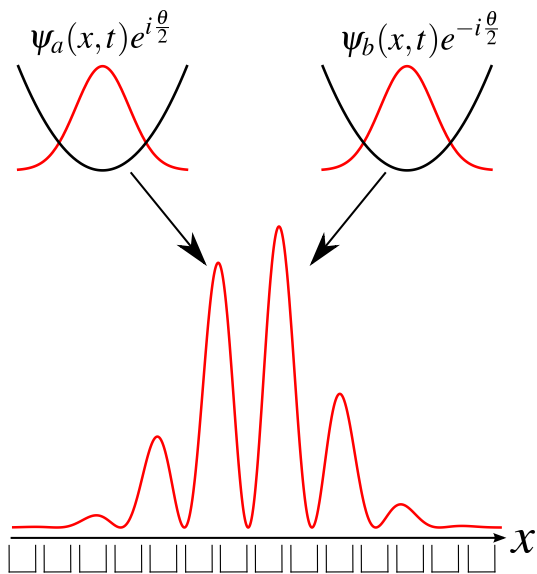}
  \caption
      {
        Schematic representation of the interferometric procedure. First, a relative phase $\theta$ is 
        imprinted between the wells. Then, the BECs are released from the trap and form an interference pattern. The 
        detectors (symbolically represented as open squares) 
        measure the positions of atoms and this data is a starting point for the phase estimation. 
      }\label{ip}
\end{figure}

The recombination of the wave-packets can be done in a simpler way, just by releasing them form the double-well trap, so they form an interference pattern, as shown in Fig.\ref{ip}.
In this manuscript we discuss how the information about the phase can be extracted from this pattern
and derive the sensitivity for different estimation strategies. 
The manuscript is organized as follows. In Section \ref{intro} we formulate the problem and introduce the basic tool -- 
the $N$-body correlation function, where $N$ is the total number of atoms. 
In Section \ref{sec_fit} we demonstrate that by performing a least-squares fit to the measured density \cite{shin}, 
the estimation sensitivity $\Delta\theta$ is bounded by the shot-noise. As discussed in detail in Section \ref{correlations}, 
in order to overcome this limit, high-order spatial correlation functions must be measured, namely, of order not smaller than $\sqrt N$. 
In particular, when estimation is performed using the $N$-th order correlation function, 
the sensitivity saturates the bound set by the Quantum Fisher Information (QFI) \cite{braun}. Then, in Section \ref{c-o-m}
we analyze an estimation scheme based on the detection of the position of the center-of-mass of the interference pattern,
which can still yield sub-shot-noise sensitivity. 
Finally, in Section \ref{mzi} we study the sensitivity of the
Mach-Zehnder interferometer implemented in a double-well, and we observe that a non-zero overlap between the wave-packets dramatically reduces the sensitivity. 
Some details of the calculations are
presented in the Appendix. The present manuscript is an extension of our previous work \cite{chwed}.

\section{The model} 
\label{intro}
To begin the discussion of different estimation methods based on position measurement,
we introduce the two-mode field operator of a bosonic gas in a
double-well potential, 
\begin{equation*}
  \hat\Psi(x,t)=\psi_a(x,t)\hat a+\psi_b(x,t)\hat b, 
\end{equation*}
where $\hat a^\dagger/\hat b^\dagger$ creates
an atom in the left/right well. With the atoms trapped, the relative phase $\theta$ is imprinted between 
the modes.
This stage is represented by a unitary evolution $\hat U(\theta)\!\!=\!\!e^{-i\theta\hat J_z}$ 
of the initial state $|\psi_{in}\rangle$ of the double-well system.
The three operators
\begin{equation*}
  \hat J_x \equiv (\hat a^\dagger\hat b^{\phantom{\dagger}}+
  \hat b^\dagger\hat a^{\phantom{\dagger}})/2,\ \ \ \hat J_y \equiv (\hat a^\dagger\hat b^{\phantom{\dagger}}-
  \hat b^\dagger\hat a^{\phantom{\dagger}})/2i\ \ \ \mathrm{and}\ \ \ \hat J_z \equiv (\hat a^\dagger\hat a^{\phantom{\dagger}}-
  \hat b^\dagger\hat b^{\phantom{\dagger}})/2
\end{equation*}
form a closed algebra of angular momentum. With the phase acquired, the trap is switched off and
the two clouds described by the mode functions $\psi_{a/b}(x,t)$ freely expand.

The most general quantity, containing the statistical information about the positions of particles forming the interference pattern, is 
the conditional probability of finding $N$ particles at positions $\vec x_N=(x_1\ldots x_N)$. It can be 
expressed in terms of the $N$-th order correlation 
function $p_N(\vec x_N|\theta)=\frac1{N!}G_N(\vec x_N,\theta)$, where
\begin{equation*}
  G_N(\vec x_N,\theta)=
  \langle\psi_{out}|
  \hat\Psi^\dagger(x_1,t)\ldots\hat\Psi^\dagger(x_N,t)\hat\Psi(x_N,t)\ldots\hat\Psi(x_1,t)|\psi_{out}\rangle.\nonumber
\end{equation*}
Here, $|\psi_{out}\rangle$ denotes the state after the interferometric transformation, 
$|\psi_{out}\rangle=e^{-i\theta\hat J_z}|\psi_{in}\rangle$.
To provide a compact and useful expression for this probability, we take following steps. 

First, we decompose the initial state in the well-population basis, 
$|\psi_{in}\rangle=\sum_{n=0}^NC_n|n,N\!\!-\!n\rangle$ and suppose
that the expansion coefficients are real and posses the symmetry $C_n=C_{N-n}$. As we will argue later, such choice of $C_n$'s is natural in context of this work. 
We switch from the Schr\"odinger to the Heisenberg representation, where the field operator evolves according to, 
\begin{equation*}
  \hat\Psi_\theta(x,t)\equiv\hat U^\dagger(\theta)\hat\Psi(x,t)\hat U(\theta)
  =\psi_a(x,t)e^{i\frac\theta2}\hat a+\psi_b(x,t)e^{-i\frac\theta2}\hat b.
\end{equation*}
The next setp is to introduce the basis of the coherent phase states \cite{laloe} defined as
\begin{equation*}
  |\varphi, N\rangle=\frac{1}{\sqrt{2^NN!}}\left(\hat a^\dagger+e^{i\varphi}\hat b^\dagger\right)^N|0\rangle
\end{equation*}
(where $|0\rangle$ is the state with zero particles). The action of the field operator on these states can be written in a simple form,
\begin{equation*}
  \hat\Psi_\theta(x,t)|\varphi, N\rangle=\sqrt{\frac N2}u_\theta(x,\varphi; t)|\varphi, N-1\rangle,
\end{equation*}
where $u_\theta(x,\varphi; t)=\psi_a(x,t)e^{\frac i2(\theta+\varphi)}+\psi_b(x,t)e^{-\frac i2(\theta+\varphi)}$.
Next, we expand the Fock states in the basis of the coherent states
\begin{equation*}
  |n,N-n\rangle=\sqrt{2^N}\frac{1}{\sqrt{{N\choose n}}}\int_0^{2\pi}\frac{d\varphi}{2\pi}
  e^{-i\varphi(N-n)}|\varphi, N\rangle.
\end{equation*}
Thus we can easily write the result of action of the field operator on the input state,
\begin{equation}\label{action}
  \fl\hat\Psi_\theta(x,t)|\psi_{in}\rangle=\sqrt{\frac{2^NN}2}
  \sum_{n=0}^N\frac{C_n}{\sqrt{N\choose n}}\int_0^{2\pi}\frac{d\varphi}{2\pi}
  e^{-i\varphi(N-n)}u_\theta(x,\varphi; t)|\varphi, N-1\rangle.
\end{equation}
Now evaluation of $G_N(\vec x_N,\theta)$ (and therefore the $p_N(\vec x_N,\theta)$ as well) is straightforward -- we act $N$ times with $\hat\Psi_\theta$ on the input state
and calculate the modulus square of the result. After normalization we obtain
\begin{eqnarray}\label{prob_ph}
  p_N(\vec x_N|\theta)&=&
  \int\limits_0^{2\pi}\int\limits_0^{2\pi}\frac{d\varphi}{2\pi}\frac{d\varphi'}{2\pi}\prod_{i=1}^Nu^*_\theta(x_i,\varphi;t)
  u_\theta(x_i,\varphi';t)\nonumber\\
  &\times&\sum_{n,m=0}^N\frac{C_nC_m\cos\left[\varphi\left(\frac N2-n\right)\right]
    \cos\left[\varphi'\left(\frac N2-m\right)\right]}{\sqrt{{N\choose n}{N\choose m}}}.
\end{eqnarray}
In the remaining part of the manuscript, we fix $t$ large enough so that the interference
pattern is formed. In this regime, the physical properties of the system change only by scaling $\sim\sqrt t$ of the characterisitc dimensions of the system.
The probability (\ref{prob_ph}) is the starting point for the following discussion of various phase estimation strategies.

\section{Estimation via the fit to the density}
\label{sec_fit}

The simplest way of estimating the value of $\theta$ is by fitting  
the average density to the interference pattern, as the position of the maximum depends on the relative phase between the two clouds. 
Such fit is commonly employed with BECs in double-wells, in order to determine, for instance, the phase coherence in the system \cite{hinds,esteve,shin}. 

In the experimental realization, the interference pattern is sampled using $M$ bins located at positions $x_i$ ($i=1\ldots M$). 
The number of particles $n_i$ in each bin 
is measured $m$ times, giving the set $n_i^{(k)}\!\!,$ $k=1,...,m$. The average occupation $\langle n_i\rangle=\lim_{m\to\infty}\sum_{k=1}^mn_i^{(k)}/m$ with free parameter $\theta$ is then fitted to the histogram
of the measured density $\{x_i, \bar{n}_i\}, i=1,...,M$, where $\bar{n}_i=\sum_{k=1}^mn_i^{(k)}/m$. 
If the size $\Delta x$ of each of $M$ bins is small, $\langle n_i\rangle$
is given by the average density
\begin{equation}\label{avgden}
 \langle n_i\rangle=G_1(x_i,\theta)\Delta x.
\end{equation}
The value of $\theta$ is determined from the least squares formula 
\begin{equation}\label{leastsq}
 \frac{d}{d\theta}\sum_{i=1}^M\frac{(\bar{n}_i-\langle n_i\rangle)^2}{2\Delta^2 n_i/m}=0.  
\end{equation} 
The fluctuations in each bin, $\Delta^2 n_i=\lim_{m\to\infty}\sum_{k=1}^m(n_i^{(k)}-\langle n_i\rangle)^2$, 
are calculated from the probability $p(n_i|\theta)$ of detecting $n_i$ particles in the $i$-th bin (for details of derrivation, see \ref{biflu}),
\begin{equation}\label{prob_bin}
  p(n_i|\theta)={N\choose {n_i}}\int\limits_{\Delta x_i}\!\!d\vec x_{n_i}\!\!\!\!\!
  \int\limits_{\mathbb{R}-\Delta x_i}\!\!\!\!d\vec x_{N-n_i}\ p_N(\vec x_N|\theta).  
\end{equation}
Using Eq.(\ref{prob_ph}) we obtain
\begin{equation}\label{fluctuations}
  \Delta^2n_i=G_1(x_i,\theta)\Delta x+\left[G_2(x_i,x_i,\theta)-G^2_1(x_i,\theta)\right](\Delta x)^2.
\end{equation}
In Eq.~(\ref{leastsq}), $\langle n_i\rangle$ and $\Delta^2 n_i$ are assumed to be known, since in the {\it phase estimation stage} the only measured data are $\bar{n}_i$. 
The quantities $\langle n_i\rangle$ and $\Delta^2 n_i$ are instead constructed during the {\it calibration stage}, preceding the phase estimation stage, 
by repeating the experiment with different known values of $\theta$. If the number of experiments in the calibration is large, 
and in absence of thermal and technical noise, the measured $\langle n_i\rangle$ and $\Delta^2 n_i$ will tend to the theoretical predictions 
given in Eq.~(\ref{avgden}) and (\ref{fluctuations}), respectively.

Our goal at this point is to determine how the quantum fluctuations $\Delta^2n_i$ in the $i$-th bin influence the sensitivity of the phase estimation via the fit (\ref{leastsq}). 
To this end, we employ the concept of the maximum likelihood estimation (MLE) \cite{crlb1,crlb2}.
\begin{figure}
  \includegraphics[scale=.36,clip]{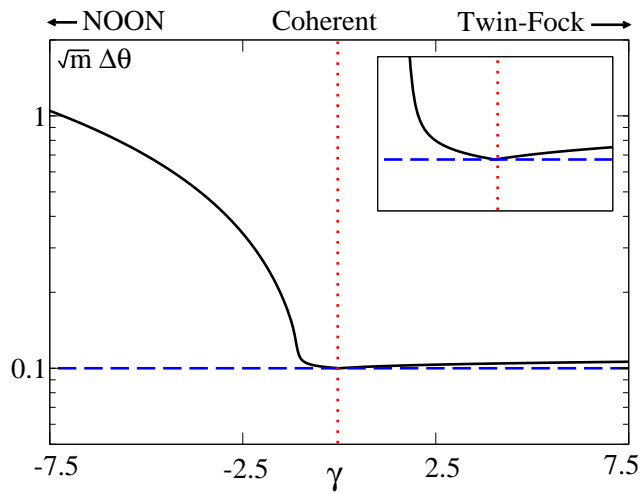}
  \caption
      {
        Sensitivity of the phase estimation from the fit to the density,
        as a function of $|\psi_{in}\rangle\in\mathcal{A}$ (solid black line) plotted using 
        $\sqrt m\Delta\theta$ with Eq.(\ref{Fa}) and $N=100$ particles.
        The blue dashed
        line represents the shot-noise limit. The horizontal dotted red line 
        indicates the position of the coherent state. 
        Clearly, the sensitivity is bounded by the shot-noise. The inset shows the
        behavior of the sensitivity in the vicinity of the coherent state.
      }\label{fit}
\end{figure}
If some quantity $\xi$ is measured, the MLE is defined as the choice of $\theta$
which maximizes the conditional probability $P(\xi|\theta)$ for the occurrence of $\xi$ given $\theta$. That is, the phase $\theta$ 
is estimated from the condition $\frac d{d\theta}P(\xi|\theta)=0$.
In case of the fit discussed here, the estimation is based on the measured average occupations $\bar{n}_i$. 
If the number of measurements $m$ is large, then according to the central limit theorem the
probability distribution for the average $\bar{n}_i$ in the $i$-th bin tends to the Gaussian
$p(\bar{n}_i|\theta)=\frac{1}{\sqrt{2\pi}\Delta n_i/\sqrt m}e^{-\frac{(\bar{n}_i-\langle n_i\rangle)^2}{2\Delta^2 n_i/m}}$. 
In every shot the atom counts are correlated between the bins. 
However, in order to link the MLE with the sensitivity of the least squares fit, we construct the likelihood function as if the measurement results 
$\bar{n}_i$ and $\bar{n}_j$, with $i\neq j$, were uncorrelated.
Thus the total probability of measuring the values 
$\{\bar{n}\}=(\bar{n}_1\ldots \bar{n}_M)$ is a product 
$P(\{\bar{n}\}|\theta)=\prod_{i=1}^Mp(\bar{n}_i|\theta)$. We note that in this case, indeed, the condition for the MLE, 
$\frac{d}{d\theta}P(\{\bar{n}\}|\theta)=0$ coincides with Eq.(\ref{leastsq}). 
The MLE sensitivity saturates the Cramer-Rao Lower Bound \cite{crlb1,crlb2}, $\Delta^2\theta=F^{-1}$. Here $F$ is the Fisher information (FI),
\begin{eqnarray}\label{ml}
  F=\sum_{\bar n_1\ldots \bar n_M=0}^N\frac{1}{P(\{\bar{n}\}|\theta)}
  \left(\frac{\partial}{\partial\theta}P(\{\bar{n}\}|\theta)\right)^2
  \rightarrow_{m\gg1}m\sum_{i=1}^M\frac{1}{\Delta^2n_i}\left(\frac{\partial\langle n_i\rangle}{\partial\theta}\right)^2.
\end{eqnarray}
Therefore, the sensitivity for the least squares fit is given by Eq.(\ref{ml}) as well. 

In the following, we demonstrate that this sensitivity is bounded by
the shot-noise. Let us assume for the moment 
that the second term in the Eq.(\ref{fluctuations}) -- which is proportional to $(\Delta x)^2$ -- can be neglected. 
Then, as can be seen from Eq.(\ref{fluctuations}),  the particle number distribution is Poissonian. The FI from (\ref{ml}) reads
\begin{equation}\label{fish}
  \fl F=m\sum_{i=1}^M\frac{1}{G_1(x_i,\theta)}\left(\frac{\partial}
  {\partial\theta}G_1(x_i,\theta)\right)^2\Delta x
  \simeq mN\!\!\!\int\limits_{-\infty}^\infty\! dx\frac{1}{p_1(x|\theta)}\left(\frac{\partial}
  {\partial\theta}p_1(x|\theta)\right)^2,
\end{equation}
with the one-particle probability 
\begin{equation}
  p_1(x|\theta)=\frac{1}{2}(|\psi_a(x,t)|^2+|\psi_b(x,t)|^2)
  +\frac2N\langle\hat J_x\rangle\mathrm{Re}\left[\psi_a^*(x,t)\psi_b(x,t)e^{i\theta}\right]. 
\end{equation}

We now calculate the Fisher information (\ref{fish}) explicitly.
As the interference pattern is formed after long expansion time,
the mode functions can be written as
\begin{equation}
  \psi_{a/b}(x,t)\simeq e^{i\frac{x^2}{2\tilde\sigma^2}\mp i\frac{x\cdot x_0}{\tilde\sigma^2}}\cdot
  \tilde\psi\left(\frac x{\tilde\sigma^2}\right),\label{long_exp}
\end{equation}
where $\tilde\sigma=\sqrt{\frac{\hbar t}{m}}$, $\tilde\psi$ is a Fourier transform of the initial wave-packets, common
for $\psi_a$ and $\psi_b$ and the separation of the wells is $2x_0$. This gives
\begin{equation*}
  F=mN\int_{-\infty}^\infty dx\big|\tilde\psi\left(\frac x{\tilde\sigma^2}\right)\big|^2
  \frac{a^2\sin^2\varphi}{1+a\cos\varphi},
\end{equation*}
with $a=\frac2N\langle\hat J_x\rangle$ and $\varphi=2\frac{x\cdot x_0}{\tilde\sigma^2}+\theta$. Notice that when the
expansion time is long, the function $\tilde\psi$ varies slowly, as compared to
the period of oscillations of $\sin\varphi$ and $\cos\varphi$. Therefore, in the above expression, one can substitute the
oscillatory part with its average value. Using the normalization of $\tilde\psi$ we obtain
\begin{equation}\label{Fa}
  F=mN\frac{a^2}2\frac{1-(-1+\sqrt{1-a^2})^2}{1+a(-1+\sqrt{1-a^2})}
\end{equation}
As $0\leq a\leq1$, we have $0\leq F\leq mN$. Therefore the Fisher information for the fit is always smaller than the shot-noise, giving  $\Delta\theta\geq\Delta\theta_{SN}=\frac{1}{\sqrt{m}}\frac{1}{\sqrt{N}}$ for any two-mode input state 
($\Delta\theta_{SN}$ denotes the shot-noise sensitivity). 
Below we argue that the inclusion of the second term in the fluctuations in Eq.~(\ref{fluctuations}) does not improve the sensitivity. 

In Eq.(\ref{fluctuations}), the first term  $G_1(x_i,\theta)\Delta x$ scales linearly with $N$ while
the second term, as a function of $N$, is a polynomial of the order not higher than two,
$aN^2+bN+c$. The fluctuations $\Delta^2n_i$ must be positive, thus $a\geq0$. 
Otherwise, for large $N$, no matter how small $\Delta x$, we would have $\Delta^2n_i<0$. 
If $a>0$, the positive second term enlarges the fluctuations and thus worsens the sensitivity. 
When $a=0$, the 
first and second terms in Eq.(\ref{fluctuations}) scale linearly with $N$, 
and for small $\Delta x$ the second term can be neglected, thus
we end up again with Eq.(\ref{fish})\footnote{ 
A similar argument shows that increasing $\Delta x$ also worsens the sensitivity with respect to Eq.(\ref{fish})}.

In order to calculate the sensitivity $\Delta\theta=\frac1{\sqrt F}$ in Eq.(\ref{Fa}),
we consider the ground states
of the BEC in a double well potential. In the two-mode approximation, the Hamiltonian of the system reads
\begin{equation}\label{ham}
  \hat H=-E_J\hat J_x+\frac{E_C}{N}\hat J_z^2.
\end{equation}
We construct a family of states $\mathcal{A}$ by finding ground states of the above Hamiltonian for various values of the ratio 
$\gamma=\frac{E_C}{NE_J}$. And so, for $\gamma>0$, the elements of $\mathcal{A}$ are number-squeezed states and tend to the twin-Fock state
$|\psi_{in}\rangle=\big|\frac N2,\frac N2\big\rangle$ with $\gamma\rightarrow\infty$.
For $\gamma<0$ the elements of $\mathcal{A}$ are phase-squeezed states \cite{grond1}. With $\gamma\rightarrow-\infty$, the ground state of (\ref{ham}) tends to the NOON state 
$|\psi_{in}\rangle=\frac1{\sqrt2}(|N0\rangle+|0N\rangle)$. 
With $\gamma=0$ we have a coherent state, $|\psi_{in}\rangle=\frac1{\sqrt{N!}}\left(\frac{\hat a^\dagger+\hat b^\dagger}{\sqrt 2}\right)^N|0\rangle$.
Notice that for all $|\psi_{in}\rangle\in\mathcal{A}$, the coefficients $C_n$, which were introduced in previous section, are real and symmetric.
For each state in $\mathcal{A}$, we calculate $a=\frac2N\langle\hat J_x\rangle$, and insert it into Eq.(\ref{Fa}). 
The sensitivity shown in Fig.\ref{fit} is clearly bounded by the shot noise.

This limitation for the sensitivity
can be explained as follows. The value of the
FI given by Eq.(\ref{fish})
is expressed in terms of the single-particle probability. We expect the useful non-classical many body correlations to decrease the value of $\Delta\theta$, but
the FI (\ref{fish}) is insensitive to these correlations, and thus must be bounded by the shot-noise. In the next section we demonstrate 
that the estimation based on the measurement of position correlations can improve the phase sensitivity.

\section{Estimation via the correlation functions}
\label{correlations}

In the estimation protocol discussed in this section, the phase $\theta$ is deduced from the measurement of the
$k$-th order correlation function 
$G_k(\vec x_k|\theta)$. As previously, we choose to deduce $\theta$ using the MLE:
a set of $k$ positions $\vec x_k$ is measured, and the phase is chosen from the condition $\frac d{d\theta}G_k(\vec x_k|\theta)=0$.
After $m\gg1$ experiments, $\Delta^2\theta=F_{(k)}^{-1}$, where 
\begin{equation}\label{fish_m}
  F_{(k)}=m\frac Nk\int\limits_{-\infty}^\infty\! d\vec x_k\frac{1}
  {p_k(\vec x_k|\theta)}\left(\frac{\partial}{\partial\theta}p_k(\vec x_k|\theta)\right)^2,
\end{equation}
with $p_k(\vec x_k|\theta)=\frac{(N-k)!}{N!}G_k(\vec x_k,\theta)$. The coefficient $\frac{N}{k}$ stands for the number
of independent drawings of $k$ particles from $N$, i.e. ${N\choose k}/{{N-1}\choose{k-1}}$. We notice 
that by setting $k=1$, i.e. the estimator is a single-particle density,
we recover the FI from Eq.(\ref{fish}). Therefore, the
measurement of positions of $N$ particles used as independent is, in terms of sensitivity, 
equivalent to fitting the average density to the interference pattern, and is limited by the shot-noise.

\begin{figure}
  \includegraphics[scale=.36,clip]{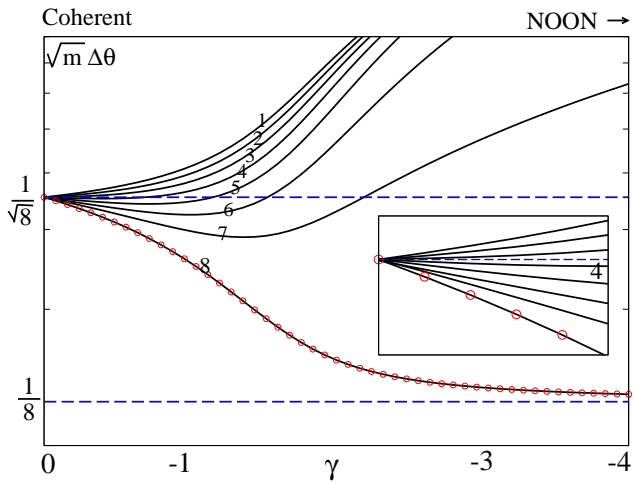}
  \caption
      {
        The sensitivity $\sqrt m\Delta\theta$ (black solid lines) for $N=8$
        calculated by numerical integration of the Eq.(\ref{fish_m}) for various $k$ as a function of 
        $|\psi_{in}\rangle\in\mathcal{A}$ with $\gamma<0$. The two limits,
        $\sqrt m\Delta\theta_{SN}$ and $\sqrt m\Delta\theta_{HL}$ are
        denoted by the upper and lower dashed blue lines, respectively. The optimal sensitivity, given by the
        QFI, is drawn with red open circles. 
        The inset magnifies the vicinity of the coherent state, showing that the sub-shot-noise sensitivity is reached 
        starting from $k_{\mathrm{min}}=4$. 
      }
      \label{corr}
\end{figure}

Let us calculate the FI for the case $k=N$, corresponding to the measurement of the full $N$-body correlation function. We represent the mode functions $\psi_{a/b}$ using 
Eq.(\ref{long_exp}).  
This expression allows to calculate $u_\theta(x,\varphi)$, and, in turn, the probability (\ref{prob_ph}), which is then inserted into 
Eq.(\ref{fish_m}). The integrals over space can be performed analytically (see \ref{ap_qfi} for details) 
and the outcome is
\begin{equation}\label{fish_q}
  F_{(k=N)}=m\cdot4\sum_{n=0}^NC_n^2\left(n-\frac N2\right)^2=m\cdot4\Delta^2\hat J_z=F_Q.
\end{equation}
Here, by $F_Q$ we denote the QFI, which is a maximal value of the Fisher information with respect to all
possible measurements \cite{braun}. In the case of pure
states, the value of the QFI is given by $4m$ times the variance of the phase-shift generator, thus in our case it reads $F_Q=m\cdot4\Delta^2\hat J_z$. 
As denoted by open circles in Fig.\ref{corr}, the Eq.(\ref{fish_q})
gives $\Delta\theta=\sqrt m\Delta\theta_{SN}$ for the coherent state ($\gamma=0$), and overcomes
this bound for all $|\psi_{in}\rangle\in\mathcal{A}$ with $\gamma<0$. The NOON state 
gives the Heisenberg limit, $\Delta\theta_{HL}=\frac1{\sqrt m}\frac1N$. 

We now discuss how the sensitivity given by the inverse of Eq.(\ref{fish_m}) changes for $k<N$. 
The space integrals for $k\neq 1,N$ cannot be evaluated analytically, thus we calculate the FI numerically taking Gaussian wave-packets
\begin{equation}\label{gauss}
      \tilde\psi\left(\frac{x}{\tilde\sigma^2}\right)=
      \left(\frac{2\sigma_0^2}{\pi\tilde\sigma^4}\right)
      ^{\frac14}e^{-\frac{x^2\sigma_0^2}{\tilde\sigma^4}}
\end{equation}
with the initial width $\sigma_0=0.1$ and 
half of the well separation $x_0=1$. 
The Fig.\ref{corr} shows how the sensitivity from Eq.(\ref{fish_m}) for $N=8$ atoms changes with increasing $k$ as a function of $|\psi_{in}\rangle\in\mathcal{A}$. 
The sensitivity improves with growing $k$, and goes below the shot-noise limit at $k_{\mathrm{min}}=4$.

For higher numbers of particles, we numerically checked that $k_{\mathrm{min}}$ tends to $\sqrt N$. Therefore, one would have to measure the correlation function of the order of
at least $\sqrt N$ in order to beat the shot-noise limit. In a realistic experiment with cold atoms, where $N\simeq 1000$, it would be very difficult to use the correlation function 
of such high order for phase estimation. The biggest difficulty resides in the calibration stage, during which one would need to experimentally probe 
a function of a $k$ dimensional variable $\vec x_k$ (with $k>\sqrt{N}$) for different values of theta.

In the following section we present a phase estimation scheme based on the measurement of the center-of-mass of the interference pattern. Although the probability for measuring
the center-of-mass at position $x$ is a function of just a one-dimensional variable, it can still provide the sub-shot-noise sensitivity. 
Nevertheless, we will demonstrate that the implementation of this estimation protocol can be challenging.

\section{Estimation via the center-of-mass measurement}
\label{c-o-m}

\subsection{Measurement of all $N$ atoms}
\label{com}

In order to estimate $\theta$ from the measurement of the center-of-mass, one has to go through a relatively simple
calibration stage.
Positions of $N$ atoms are recorded independently and from this data location of the center-of-mass is deduced.
Many repetitions of the experiment give the function $p_{cm}(x|\theta)$ of a one-dimensional variable. 
The expression for this function can be extracted from the full $N$-body probability (\ref{prob_ph}) by
\begin{equation*}\label{pcm}
  p_{cm}(x|\theta)=\int d\vec x_N\,\delta\left(x-\frac1N\sum_{i=1}^Nx_i\right)p_N(\vec x_N|\theta),
\end{equation*}
where ``$\delta$'' is the Dirac delta. 
To provide an analytical expression for this probability, we model the mode-functions by Gaussians as in Eq.(\ref{gauss}). Using a reasonable assumption that the initial
separation of the wave-packets is much larger than their width, i.e. $e^{-x_0^2/\sigma_0^2}\ll1$, we obtain
\begin{equation}\label{prob_cm}
  \fl p_{cm}(x|\theta)=\sqrt{\frac{2\sigma_0^2N}{\pi\tilde\sigma^4}}e^{-\frac{2x^2\sigma^2_0}{\tilde\sigma^4}N}
  \left[1+\frac12(C_0+C_N)^2\cos\left(N\theta+\frac{2 N x_0}{\tilde\sigma^2}x\right)\right].
\end{equation}
The details of this derivation are presented in \ref{app_cm}.
Notice an interesting property -- the above probability
depends on $\theta$ only for states with non-negligible NOON components $C_0$ and $C_N$, as already noticed in \cite{bach}. 

\begin{figure}
  \includegraphics[scale=.36,clip]{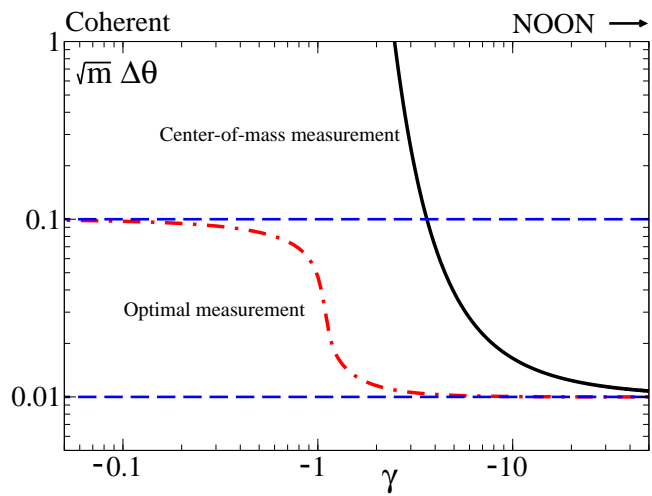}
  \caption
      {
        The sensitivity $\sqrt m\Delta\theta$ (black solid line)  for $N=100$
        calculated with Eq.(\ref{fish_cm}) as a function of $|\psi_{in}\rangle\in\mathcal{A}$ with $\gamma<0$. 
        The values of $\sqrt m\Delta\theta_{SN}$ and $\sqrt m\Delta\theta_{HL}$ are
        denoted by the upper and lower dashed blue lines, respectively. The optimal sensitivity, given by the inverse of
        the QFI, is drawn with the red open circles. 
      }
      \label{cm}
\end{figure}

When $p_{cm}(x|\theta)$ is known,
the phase can be estimated using the MLE, as used in the previous sections. Then once again 
the sensitivity is given by the inverse of the FI, which can be calculated analytically,
\begin{equation}\label{fish_cm}
  \fl F_{cm}=m\int\limits_{-\infty}^\infty\frac{dx}{p_{cm}(x|\theta)}\left(\frac{\partial}{\partial\theta}p_{cm}(x|\theta)\right)^2=mN^2\left[1-\sqrt{1-\frac12(C_0+C_N)^2}\right],
\end{equation}
where $m$ is the number of experiments. 
In Fig.\ref{cm} we plot the sensitivity calculated by the inverse of the FI (\ref{fish_cm}) as
a function of $|\psi_{in}\rangle\in\mathcal{A}$ with $\gamma\leq0$.
Although the estimation through the center-of-mass is not optimal ($\Delta\theta>\frac1{\sqrt{F_Q}}$), the sensitivity can be better than the shot-noise,
with $\Delta\theta\rightarrow\Delta\theta_{HL}$ for $|\psi_{in}\rangle\rightarrow$ NOON.
The calibration stage is not as difficult as in the case of high-order correlations, however
phase estimation based on the center-of-mass measurement demands detection of all $N$ atoms \cite{dobrzanski1, dobrzanski2, dobrzanski3,knysh}, as we show below.

\subsection{Measurement of $k<N$ atoms}
\begin{figure}
  \includegraphics[scale=.36,clip]{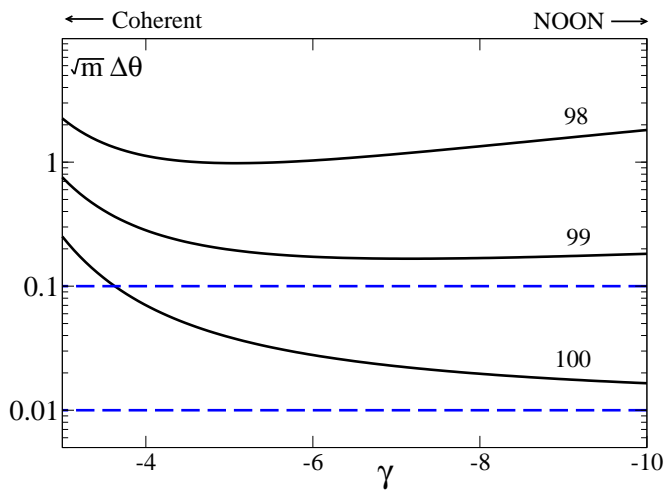}
  \caption
      {
        The sensitivity $\sqrt m\Delta\theta$ (black solid line)  for $N=100$
        calculated with Eq.(\ref{fk}) as a function of $|\psi_{in}\rangle\in\mathcal{A}$ with $\gamma<0$. 
	The values of $\sqrt m\Delta\theta_{SN}$ and $\sqrt m\Delta\theta_{HL}$ are
        denoted by respectively the upper and the lower dashed blue line. The three solid lines correspond to the
	phase sensitivity for estimation of the center-of-mass with different number of particles. For $k=100$,
	the sensitivity is below the shot-noise and tends to $\sqrt m\Delta\theta_{HL}$ for $|\psi_{in}\rangle\rightarrow$
	NOON. As soon as $k\neq N$, the sub-shot-noise sensitivity is lost and the value of $\sqrt m\Delta\theta$ 
	increases dramatically.
      }
      \label{noon}
\end{figure}

If the measurement of the center-of-mass is based on detection of $k<N$ atoms, the probability (\ref{prob_cm}) transforms into
\begin{equation}\label{prob_cm_k}
  p^{(k)}_{cm}(x|\theta)=\int d\vec x_k\,\delta\left(x-\frac1k\sum_{i=1}^kx_i\right)p_k(\vec x_k|\theta),
\end{equation} 
where $p_k(\vec x_k|\theta)=\int d\vec x_{N-k} p_N(\vec x_N|\theta)$. The probability (\ref{prob_cm_k}) can be calculated
in a manner similar to that presented in \ref{app_cm}. The result is
\begin{equation}\label{pk_fin}
  p^{(k)}_{cm}(x|\theta)=\sqrt{\frac{2\sigma_0^2k}{\pi\tilde\sigma^4}}e^{-\frac{2x^2\sigma^2_0}{\tilde\sigma^4}k}
  \left[1+a\cos\left(k\theta+\frac{2k x_0}{\tilde\sigma^2}x\right)\right],
\end{equation}
where
\begin{equation}
  a=2\sum_{i=0}^{N-k}{(N-k)\choose i}\frac{C_iC_{i+k}}{\sqrt{{N\choose{i+k}}{N\choose i}}}.\label{a}
\end{equation}
Notice that for $k=N$ we recover the result from the previous section $a=2C_0C_N=\frac12(C_0+C_N)^2$ 
(as we are using the symmetric states, $C_0=C_N$). The FI for the probability (\ref{pk_fin}) can be calculated analytically,
\begin{equation}\label{fk}
  F=mk^2(1-\sqrt{1-a^2}).
\end{equation}
Let us now evaluate $a$ -- and thus $F$ -- for various $k\simeq N$. For $k=N$ and the NOON state, we have 
$C_0=C_N=\frac1{\sqrt2}$, giving $a=1$ and $F=mN^2$. From Eq.(\ref{a}) we notice that, for any $k$, 
$a$ is the sum of $N-k$ terms, each depending on the coefficients $C_i$ and $C_{i+k}$. And so, for $k=N-1$, $a$ will be maximal for a NOON-like state with $C_0=C_{N-1}=\frac12$ and $C_1=C_N=\frac12$. For this state we obtain $a=\frac1{\sqrt N}$, and for large $N$ the value of the FI is $F=mN$. Therefore, the phase estimation using the center-of-mass of $N-1$ particles gives a sensitivity bounded by the shot-noise. 
Each loss of an atom decreases the FI roughly by a
factor of $N$, drastically deteriorating the sensitivity.
In Fig.\ref{noon} we plot the sensitivity $\sqrt m\Delta\theta$ calculated with the FI
from Eq.(\ref{fk}) for various $k\simeq N$. To calculate $a$, we choose the subset of $|\psi_{in}\rangle\in\mathcal{A}$
which are in the vicinity of the NOON state. The Figure shows a dramatic loss of sensitivity as soon as $k\neq N$.

\section{Estimation via the position measurement with the Mach-Zehnder Interferometer}
\label{mzi}

\subsection{Formulation}
\label{mzi_gen}

So far, we focused on the position
measurement of atoms released from a double-well trap, and studied the phase estimation sensitivity.
The fit to the density gives sensitivity limited by the shot-noise, and
this bound can be overcome by phase estimation with correlation functions of the order of at least $\sqrt N$.
As it is difficult to measure these correlations in the experiment, it will be challenging to beat the shot-noise 
limit using the interference pattern.
Although the sensitivity of the phase estimation based on the center-of-mass measurement can also be sub-shot-noise,
the protocol is extremely vulnerable to the loss of particles.

In the above scenario, the sub-shot-noise sensitivity, which relies on non-classical {\it particle} correlations, 
is reached by directly measuring spatial correlations between the atoms forming the interference pattern and using the 
latter as estimators for the phase shift. On the other hand, it is well known that the Mach-Zehdner Interferometer (MZI)
can easily provide sub-shot-noise sensitivity just by a simple measurement of the population imbalance between 
the two arms and a proper choice of the input state $|\psi_{in}\rangle$. 
This is because, in the MZI, the correlations between the two modes carry the part of the information contained 
in the particle correlations which is useful for phase estimation. When the clouds are released 
from the trap and the two modes start to overlap, the correlations between the two modes are lost,
since an atom detected in the overlap region cannot be told to have come from either of the two initially separated clouds.
This is the reason why it is necessary then to measure directly high-order spatial correlation functions in 
order to reach sub-shot-noise sensitivity.

It would be thus interesting to quantify the effect of the wave-packets' overlap on the sensitivity of the MZI. 
This analysis has also a practical interest since, in the implementation of the atomic MZI, 
the precision of the  population imbalance measurement can be improved by opening the trap and letting
the clouds expand for a while. In this way the density of the clouds drops, facilitating the measurement 
of the number of particles. However, during the expansion, the clouds inevitably start to overlap, leading to loss of
information about the origin of the particles, as noted above. 
In this section, we show how the increasing overlap deteriorates the sensitivity of the MZI in two estimation scenarios.

The MZI consists of three stages: two beam-splitters represented by unitary evolution operators $e^{\mp i\frac\pi2\hat J_x}$
separated by the phase acquisition $e^{-i\theta\hat J_z}$. 
The atomic MZI can be realized as follows. Consider a two-mode system governed by the Hamiltonian (\ref{ham}) with $E_C=0$.
The first beam-splitter is done by letting the atoms tunnel between the two wells for $t=\frac\pi2\frac\hbar{E_J}$ so
the unitary evolution operator reads $\hat U_1=e^{-i\frac\pi2\hat J_x}$. Then, an inter-well barrier is raised,
in order to supress the oscillations ($E_J=0$) and a phase between the wells is imprinted, giving 
$\hat U_2=e^{-i\theta\hat J_z}$. The interferometric sequence
is closed by another beam-splitter, $\hat U_3=e^{i\frac\pi2\hat J_x}$. The full evolution operator reads
\begin{equation*}
  \hat U(\theta)=\hat U_3\hat U_2\hat U_1=e^{i\frac\pi2\hat J_x}e^{-i\theta\hat J_z}e^{-i\frac\pi2\hat J_x}=
  e^{-i\theta\hat J_y},
\end{equation*}
where we used commutation relations of the angular momentum operators. 

In order to analyze the sensitivity of the MZI,
we introduce the conditional probability $p_N(\vec x_N|\theta)$ of detecting $N$ atoms at positions 
$\vec x_N=x_1\ldots x_N$. To evaluate this probability for any initial state of the double-well system 
$|\psi_{in}\rangle$, we take the same steps as in Section II. In the Heisenberg representation, the field operator
evolves as
\begin{eqnarray}\label{field_mzi}
  &&\hat\Psi_\theta(x,t)\equiv\hat U^\dagger(\theta)\hat\Psi(x,t)\hat U(\theta)\\
  &&=\left[\psi_a(x,t)\cos\left(\frac\theta2\right)+\psi_b(x,t)\sin\left(\frac\theta2\right)\right]\hat a\nonumber\\
  &&+\left[\psi_b(x,t)\cos\left(\frac\theta2\right)-\psi_a(x,t)\sin\left(\frac\theta2\right)\right]\hat b\nonumber.
\end{eqnarray}
Again, we express the action of the field operator on $|\psi_{in}\rangle$ using the basis of the coherent phase-states
and obtain Eq.(\ref{action}) with
\begin{eqnarray*}
  &&u_\theta(x,\varphi;t)=\\
  &&=\left[\psi_a(x,t)\cos\left(\frac\theta2\right)+\psi_b(x,t)\sin\left(\frac\theta2\right)\right]e^{i\frac\varphi2}\\
  &&+\left[\psi_b(x,t)\cos\left(\frac\theta2\right)-\psi_a(x,t)\sin\left(\frac\theta2\right)\right]e^{-i\frac\varphi2}.
\end{eqnarray*}
Therefore, the probability $p_N(\vec x_N|\theta)$ for the MZI is given by Eq.(\ref{prob_ph}) with the 
$u_\theta(x,\varphi;t)$ function defined above. 

\subsection{Measurement of the population imbalance}
The most common phase-estimation protocol discussed in context of the MZI is the measurement of the population imbalance
between the two arms of the interferometer. In order to assess how the sensitivity of this protocol is influenced
by the expansion of the wave-packets, we introduce the probability of measuring $n_L$ atoms in the left sub-space as follows
\begin{equation}\label{imb}
  p_{imb}(n_L|\theta)={N\choose{n_L}}
  \int\limits_{-\infty}^0\!\!d\vec x_{n_L}\!\!\int\limits_0^{\infty}\!d\vec x_{N-n_L}\,p_N(\vec x_N|\theta). 
\end{equation}
This probability depends on the expansion time via $\psi_{a,b}(x,t)$, which enter the definition of
$p_N(\vec x_N|\theta)$. 
The sensitivity for various expansion times, if $m\gg1$ measurements are performed, can be calculated using the error propagation formula,
\begin{equation}\label{sens_imb}
  \Delta^2\theta=\frac1m\frac{\Delta^2n}{\Big|\frac{\partial\langle n\rangle}{\partial\theta}\Big|^2},
\end{equation}
where 
\begin{equation*}
  \langle n\rangle=\sum_{n_L=0}^Np_{imb}(n_L|\theta)\left(n_L-\frac N2\right)
\end{equation*}
is the average value of the population imbalance and
\begin{equation*}
  \Delta^2n=\sum_{n_L=0}^Np_{imb}(n_L|\theta)\left(n_L-\frac N2\right)^2-\langle n\rangle^2
\end{equation*}
are the associated fluctuations. The probability (\ref{imb}) resembles
Eq.(\ref{prob_bin}), and the above moments are calculated as in \ref{biflu} resulting in
\begin{equation*}
  \fl\langle n\rangle=\int\limits_0^\infty\!dx\,G_1(x|\theta)-\frac N2\ \ \ \ \ \mathrm{and}\ \ \ \ \ 
  \Delta^2n=\frac{N^2}4-\int\limits_0^\infty\int\limits_{-\infty}^0\!d\vec x_2\,G_2(\vec x_2|\theta)-\langle n\rangle^2.
\end{equation*}
The two lowest correlation functions for the MZI read 
$G_1(x|\theta)=\langle\hat\Psi^\dagger_\theta(x,t)\hat\Psi_\theta(x,t)\rangle$ and 
$G_2(\vec x_2|\theta)=
\langle\hat\Psi^\dagger_\theta(x_1,t)\Psi^\dagger_\theta(x_2,t)\hat\Psi_\theta(x_2,t)\hat\Psi_\theta(x_1,t)\rangle$, with
the field operator from Eq.(\ref{field_mzi}), and the averages calculated with the input state. When the
two wave-packets don't overlap, i.e. $\psi_a(x,t)\psi^*_b(x,t)\simeq0$ for all $x\in\mathbb{R}$, 
Eq.(\ref{sens_imb}) simplifies to
\begin{equation}\label{sens_ep}
  \Delta^2\theta=\frac1m\frac{\Delta^2\hat J_x\sin^2\theta+\langle\hat J_z^2\rangle\cos^2\theta}
	{\langle\hat J_x\rangle^2\cos^2\theta}.
\end{equation}
This is the well-known expression for the sensitivity of the population imbalance between separated arms.
It gives $\Delta\theta\leq\Delta\theta_{SN}$ for all $|\psi_{in}\rangle\in\mathcal{A}$ with $\gamma\geq0$.

\begin{figure}
  \includegraphics[scale=.36,clip]{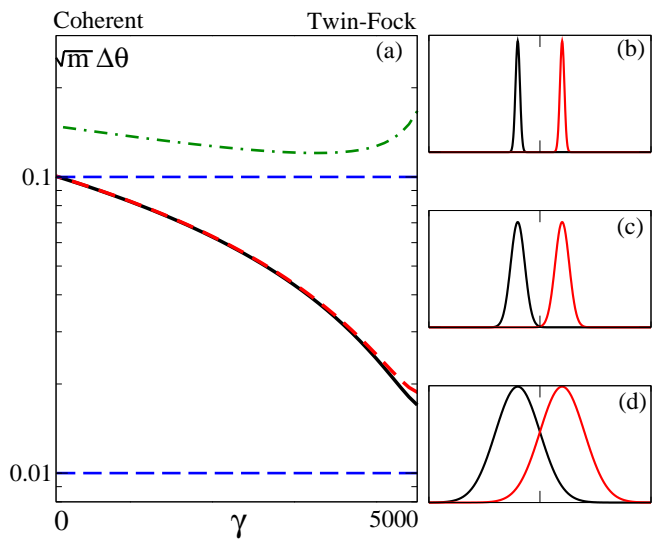}
  \caption
      {
        (a) The sensitivity $\sqrt m\Delta\theta$ calculated with Eq.(\ref{sens_imb}) for 
	three different expansion times $\tau$ as a function of $|\psi_{in}\rangle\in\mathcal{A}$ with $\gamma\geq0$.
	The solid black line corresponds to the setup shown in (b), where $\tau=0$ and the wave-packets don't overlap.
	The dashed red line corresponds to (c), where $\tau=3$ and the wave-packets start to overlap.
	The dot-dashed green line corresponds to (d), where $\tau=10$ and the wave-packets strongly overlap.
	Clearly, the sensitivity is influenced by  any non-vanishing overlap.
	The values of $\sqrt m\Delta\theta_{SN}$ and $\sqrt m\Delta\theta_{HL}$ are
        denoted by respectively the upper and the lower dashed blue line. Here, $N=100$ and $\theta=0$.
      }
      \label{expand}
\end{figure}

We investigate the impact of the overlap on the sensitivity (\ref{sens_imb}) by modelling the free expansion 
of the wave-packets $\psi_{a/b}(x,t)$ by Gaussians, 
\begin{equation*}
  \psi_{a/b}(x,t)=\frac{1}{(2\pi\sigma_0^2(1+i\tau))^{1/4}}e^{-\frac{(x\pm x_0)^2}{4\sigma^2(1+i\tau)}},
\end{equation*}
and take $x_0=1$ and the initial width $\sigma_0=0.1$. In Fig. \ref{expand} 
we plot the sensitivity $\sqrt m\Delta\theta$ taking $\theta=0$ and $N=100$ for three different expansion times 
$\tau$. The initial sensitivity deteriorates as soon as the condensates start to overlap, and the sub-shot-noise
sensitivity is lost for long expansion times. We attribute this decline to the loss of information 
about the correlations between the modes. Therefore, special attention has to be paid to avoid the overlap
when letting the two trapped 
condensates spread. Although we expect that the expansion facilitates the atom-number measurement, the conclusion of this
section is that any overlap of the spatial modes has a strong negative impact on the sensitivity of the MZI.

\subsection{Estimation via the center-of-mass measurement for the MZI}
In Section \ref{com}, we have demonstrated that when the two wave-packets overlap and form an interference pattern, phase estimation based on the 
center-of-mass measurement can give sub-shot-noise
sensitivity. Here we study the same estimation strategy applied in the MZI case.
Again, we start with the probability 
$p_{cm}(x|\theta)$ of measuring the center-of-mass at position $x$,
\begin{equation*}
  p_{cm}(x|\theta)=\int d\vec x_N\,\delta\left(x-\frac1N\sum_{i=1}^Nx_i\right)p_N(\vec x_N|\theta).
\end{equation*}
Using $p_N(\vec x_N|\theta)$ for the MZI, one can analytically 
calculate the center-of-mass probability only in the limit of small $\theta$,
\begin{equation}\label{cm_mzi}
  \fl p_{cm}(x|\theta)=\frac{N}{2\pi\sigma^2}\left[\sum_{l=0}^NC_l^2f_l(x)+
    \theta\sum_{l=0}^NC_lC_{l+1}\sqrt{(l+1)(N-l)}\left(f_{l+2}(x)-f_l(x)\right)\right],
\end{equation}
where
\begin{equation*}
  f_l(x)=\exp\left[-\frac{(x-x_0\left(\frac{2l}{N}-1)\right)^2}{2\sigma_0^2}\right].
\end{equation*}
With this probability, we can again calculate the sensitivity using the error propagation formula \cite{crlb1,crlb2}, 
\begin{equation*}
  \Delta^2\theta=\frac1m\frac{\Delta^2x}{\Big|\frac{\partial\langle x\rangle}{\partial\theta}\Big|^2},
\end{equation*}
where
\begin{equation*}
  \langle x\rangle=\!\!\!\int\limits_{-\infty}^\infty\!\!dx\,p_{cm}(x|\theta)x\ \ \mathrm{and}\ \ 
  \Delta^2x=\!\!\!\int\limits_{-\infty}^\infty\!\!dx\,p_{cm}(x|\theta)x^2-\langle x\rangle^2.
\end{equation*}
These two moments can be easily calculated with Eq.(\ref{cm_mzi}), giving, in the limit $\theta\rightarrow0$,
\begin{equation}\label{sens_mzi_cm}
  \Delta^2\theta\Big|_{\theta\rightarrow0}=\frac1m\left[\frac{\langle\hat J_z^2\rangle}{\langle\hat J_x\rangle^2}
    +\left(\frac{\sigma}{x_0}\right)^2\frac N{4\langle\hat J_x\rangle^2}\right].
\end{equation}
Notice that when the initial size of the Gaussians tends to zero, we recover the sensitivity from Eq.(\ref{sens_ep}) 
(in the limit of $\theta\rightarrow0$). This is not surprising, as when the mode-function are point-like, the
measurements of the center-of-mass and the measurement of the population imbalance are equivalent, and related
by $x_{cm}=2x_0\frac nN$. Therefore, for small $\sigma$, the measurement of the center-of-mass yields sub-shot-noise
sensitivity for all $|\psi_{in}\rangle\in\mathcal{A}$ with $\gamma>0$. However, for non-zero $\sigma$, the second term in
Eq.(\ref{sens_mzi_cm}) spoils the sensitivity. This is because $\frac{N}{4\langle\hat J_x\rangle^2}\geq\frac1N$ 
is always satisfied. 
Even if the first term scales better than at the shot-noise limit, the other one does not,
and will dominate for large $N$. 

>From what we presented in this Section, we conclude that both the population imbalance and the center-of-mass measurements can give sub-shot-noise sensitivity
for the MZI, but both are very sensitive to the growing size of the wave-packets.

\section{Conclusions}
In this manuscript we have discussed in detail how the measurement of positions
of atoms forming an interference pattern can be useful in context of atom interferometry. 
We showed that the phase estimation based on the fit to the density gives sensitivity limited by the shot-noise, because the FI is expressed in terms 
of the single-particle probability only. The sensitivity can be improved below the shot-noise limit by estimating the phase using correlation functions of order at least $\sqrt N$. 
Moreover, we demonstrated that the information contained in the $N$-th order correlation function allows to perform an optimal detection strategy, 
reaching Heisenberg-limited sensitivity when NOON states are used. 
We also showed that the measurement of the position of the center-of-mass of the interference pattern 
gives sub-shot-noise sensitivity for all states with non-negligible NOON component. 
Both the measurement of high-order correlations and the center-of-mass position
are difficult to perform. The former requires the construction of a function of highly-dimensional variables, 
while the latter works well only if all $N$ atoms forming the interference pattern are detected. 
We attribute the difficulty to obtain the sub-shot-noise sensitivity to the fact 
that, after formation of the interference pattern, the modes cannot be distinguished,
and the information useful for interferometry is only contained in the correlations between the particles. 
These correlations are very difficult to extract from the experimental data,
therefore reaching sub-shot-noise sensitivity with two interfering BECs might prove very challenging. 
In the final part of this work, we turned our attention to the MZI, which is known to provide sub-shot-noise sensitivity for the simpler measurement of the
population imbalance between the two clouds. We have shown that the sensitivity of the MZI is strongly influenced by a non-zero overlap between the two wave-packets, 
both in case of the population imbalance and the center-of-mass measurement.

\appendix
\section{Bin fluctuations}
\label{biflu}
To derive the expression for the average number and the fluctuations of the atom count in a bin, we use the
probability from Eq.(\ref{prob_bin}). With help of  Eq.(\ref{prob_ph})
we obtain
\begin{equation*}
  \fl p(n_i|\theta)=
  \int\limits_0^{2\pi}\int\limits_0^{2\pi}\frac{d\varphi}{2\pi}\frac{d\varphi'}{2\pi}{N\choose{n_i}}
  a_\theta^{n_i}b_\theta^{N-n_i}\sum_{n,m=0}^N\frac{C_nC_m\cos\left[\varphi\left(\frac N2-n\right)\right]
    \cos\left[\varphi'\left(\frac N2-m\right)\right]}{\sqrt{{N\choose{n}}{N\choose{m}}}}\nonumber,
\end{equation*}
where
\begin{eqnarray*}
  &&a_\theta=\int\limits_{\Delta x_i}\!\!\!dx\,
  u^*_\theta(x,\varphi;t)u_\theta(x,\varphi';t)\ \ \ \mathrm{and}\ \ \ 
  b_\theta=\int\limits_{\mathbb{R}-\Delta x_i}\!\!\!\!\!\!dx\,u^*_\theta(x,\varphi;t)u_\theta(x,\varphi';t).
\end{eqnarray*}
Since the distribution of $n_i$ inside the integrals is binomial, one can easily calculate the average value as follows
\begin{eqnarray*}
  \langle n_i\rangle&=&\sum_{n_i=0}^Nn_i\,p(n_i|\theta)=N\int\limits_0^{2\pi}
  \int\limits_0^{2\pi}\frac{d\varphi}{2\pi}\frac{d\varphi'}{2\pi}
  a_\theta(a_\theta+b_\theta)^{N-1}\nonumber\\
  &\times&\sum_{n,m=0}^N\frac{C_nC_m\cos\left[\varphi\left(\frac N2-n\right)\right]
    \cos\left[\varphi'\left(\frac N2-m\right)\right]}{\sqrt{{N\choose n}{N\choose m}}}\nonumber.
\end{eqnarray*}
Notice, that using the definition of $a_\theta$ and $b_\theta$, this can be rewritten as
\begin{equation*}
  \langle n_i\rangle=N\int\limits_{\Delta x_i}\!\!\!dx_1\!\!\!
  \int\limits_{-\infty}^\infty d\vec x_{N-1}\,p_N(\vec x_N|\theta)=
  \!\!\!\int\limits_{\Delta x_i}\!\!\!dx_1G_1(x_1|\theta),
\end{equation*}
where $G_1(x_1|\theta)=\langle\Psi^\dagger_\theta(x_1,t)\Psi_\theta(x_1,t)\rangle$ is a one-particle density.
In the same manner, one can demonstrate that the fluctuations of the number of particles in the $i$-th bin are given by
\begin{equation*}
  \Delta^2n_i=\langle n_i^2\rangle-\langle n_i\rangle^2=
  \int\limits_{\Delta x_i}\!dx_1\ G_1(x_1,\theta)+\int\limits_{\Delta x_i}\int\limits_{\Delta x_i}\!d\vec x_2\ G_2(\vec x_2,\theta)
  -\langle n_i\rangle^2.
\end{equation*}
When the bin size is much smaller than the characteristic variation of the mode functions, the above integrals
are approximated by
\begin{equation*}
  \fl\langle n_i\rangle=G_1(x_i,\theta)\Delta x\ \ \ \mathrm{and}\ \ \ \Delta^2n_i=G_1(x_i,\theta)\Delta x+\left[G_2(x_i,x_i,\theta)-G^2_1(x_i,\theta)\right](\Delta x)^2.
\end{equation*}

\section{The FI for $k=N$}
\label{ap_qfi}
In this Appendix we prove the Eq.(\ref{fish_q}). We start from the expression for the $N$-particle probability 
(\ref{prob_ph}) for the wave-packets after long expansion time, as in Eq.(\ref{long_exp}). A straightforward evaluation
shows that in this case, the probability can be written as
\begin{equation}\label{prob_app}
  p_N(\vec x_N|\theta)=\left[f(\vec x_N|\theta)\right]^2,
\end{equation}
where $f$ is real and reads
\begin{equation}\label{f2}
  \fl f(\vec x_N|\theta)=2^N\!\!\!\int_0^{2\pi}\!\!\frac{d\varphi}{2\pi}
  \sum_{n=0}^N \frac{C_n\cos\left[\varphi\left(\frac{N}{2}-n\right)\right]}{\sqrt{{N\choose{n}}}}\prod_{i=1}^N\tilde\psi\left(\frac{x_i}{\tilde\sigma^2}\right)
  \cos\left(\frac{x_0 x_i}{\tilde\sigma^2}+\frac\theta2+\frac\varphi2\right).
\end{equation}
Moreover, we assume that the initial separation of the wells of the trapping potential is much larger than the width of
the mode functions. Under this assumption, the probability (\ref{prob_app}) is normalized. This probability is inserted
into the definition of the Fisher information. As $f$ is real, we immediately obtain
\begin{equation*}
  F_{(k=N)}=m\cdot4\int d\vec x_N\left[f(\vec x_N|\theta)\right]^2.
\end{equation*}
Now, the order of the integration can be reversed, and the space integrals are performed first. As the mode-functions
are normalized, the result is following
\begin{equation*}
  \fl F_{(k=N)}=mN2^{N}\sum_{n=0}^N\frac{C_n^2}{{N\choose{n}}}
  \int_0^{2\pi}\frac{d\varphi}{2\pi}\cos\left[\varphi(N-2n)\right]\left[N\left(\cos(\varphi)\right)^{N}-(N-1)\left(\cos(\varphi)\right)^{N-2}\right].
\end{equation*}
The phase integral can be now easily evaluated, giving
\begin{equation*}
  F_{(k=N)}=m\cdot4\sum_{n=0}^NC_n^2\left(n-\frac N2\right)^2=m\cdot4\Delta^2\hat J_z,
\end{equation*}
which is the Eq.(\ref{fish_q}).

\section{Evaluation of the center-of-mass probability}
\label{app_cm}
In this Appendix we derive the expression for the probability of detecting the center-of-mass at position $x$, as in
Eq.(\ref{prob_cm}). The definition of $p_{cm}(x|\theta)$ relates it to the full $N$-body probability by
\begin{equation*}
  p_{cm}(x|\theta)=\int d\vec x_N\,\delta\left(x-\frac1N\sum_{i=1}^Nx_i\right)p_N(\vec x_N|\theta).
\end{equation*}
To calculate this probability we assume a long expansion time and use Eqs (\ref{prob_app}) and (\ref{f2}). Then, we notice
that the Dirac delta can be represented as the Fourier transform,
\begin{equation*}
  p_{cm}(x|\theta)=\frac{1}{2\pi}\int\limits_{-\infty}^\infty\!\! dk\!\!
  \int\!\! d\vec x_N\,e^{-ik(x-\frac1N\sum_{i=1}^Nx_i)}p_N(\vec x_N|\theta).
\end{equation*}
This equation can be rewritten in a following way
\begin{equation}\label{trans_app}
  p_{cm}(x|\theta)=\frac{1}{2\pi}\int\limits_{-\infty}^\infty dk\,e^{-ikx}\,\tilde p_{cm}(k|\theta)
\end{equation}
where
\begin{equation*}
  \tilde p_{cm}(k|\theta)=\int e^{i\frac kN\sum_{i=1}^Nx_i} p_N(\vec x_N|\theta)\,d\vec x_N
\end{equation*}
is  the Fourier transform of the probability $\tilde p_{cm}(x|\theta)$. To provide an analytical expression for 
this probability, we assume that the initial wave-packets are Gaussian as in Eq.(\ref{gauss}).
The integration over space is performed giving
\begin{eqnarray}\label{fou_app}
  &&\tilde p_{cm}(k|\theta)=\int_0^{2\pi}\frac{d\varphi}{2\pi}\int_0^{2\pi}\frac{d\varphi'}{2\pi}
  \left[I(k,\varphi,\varphi')\right]^N\\
  &&\times\sum_{n,m=0}^N \frac{C_nC_m}{\sqrt{N\choose n}{N\choose{m}}}\cos\left[\varphi\left(\frac{N}{2}-n\right)\right]
  \cos\left[\varphi'\left(\frac{N}{2}-m\right)\right],\nonumber
\end{eqnarray}
where
\begin{equation*}
  \fl I(k,\varphi,\varphi')=e^{i\left(\frac{\varphi+\varphi'}{2}+\theta\right)}e^{-\frac{(k+k_0)^2}{2w^2}}+
  e^{-i\left(\frac{\varphi+\varphi'}{2}+\theta\right)}e^{-\frac{(k-k_0)^2}{2w^2}}
  +2\cos\left(\frac{\varphi+\varphi'}2\right)e^{-\frac{k^2}{2w^2}},\nonumber
\end{equation*}
with $k_0=\frac{2 N x_0}{\tilde\sigma^2}$ and $w=\frac{2 N \sigma_0}{\tilde\sigma^2}$. The function 
$I(k,\varphi,\varphi')$ consists of three peaks, located at $k=\pm k_0$ and $k=0$. 
When the well separation $2x_0$ is large compared to the initial width of the trapped wave-packets $\sigma$,
then these three peaks are separated and do not overlap. 
Therefore, $\left[I(k,\varphi,\varphi')\right]^N$ can be approximated by the sum of $N$-th powers of its three components
\begin{equation*}
  \fl\left[I(k,\varphi,\varphi')\right]^N\simeq 
  e^{iN\left(\frac{\varphi+\varphi'}{2}+\theta\right)}e^{-\frac{N(k+k_0)^2}{2w^2}}+
  e^{-iN\left(\frac{\varphi+\varphi'}{2}+\theta\right)}e^{-\frac{N(k-k_0)^2}{2w^2}}
  +2^N\cos^N\left(\frac{\varphi+\varphi'}2\right)e^{-\frac{Nk^2}{2w^2}}.\nonumber
\end{equation*}
This result is put into Eq.(\ref{fou_app}), the phase integrals are evaluated and the result is
\begin{equation*}
  \tilde p_{cm}(k|\theta)=e^{-N\frac{k^2}{2w^2}}+\left[e^{iN\theta-N\frac{(k+k_0)^2}{2w^2}}+
    e^{-iN\theta-N\frac{(k-k_0)^2}{2w^2}}\right]\frac{(C_0+C_N)^2}4.
\end{equation*}
Then, using Eq.(\ref{trans_app}), we obtain
\begin{equation*}
  p_{cm}(x|\theta)=\frac{w\,e^{-\frac{w^2x^2}{2N}}}{\sqrt{2\pi N}}\left[1+\frac{(C_0+C_N)^2}2\cos(N\theta+k_0x)\right],
\end{equation*}
which, with help of definitions of $w$ and $k_0$, gives Eq.(\ref{prob_cm}).


\section*{References}

\begin{thebibliography}{40}

\bibitem{mach}{Mach E, ``The Principles of Physical Optics'', Dover (2003)}

\bibitem{cronin} {Cronin A D, Schmiedmayer J and Pritchard D E, \RMP {\bf 81}, 1051 (2009)}

\bibitem{cornell}{Obrecht J M, Wild R J, Antezza M, Pitaevskii L P, Stringari S and Cornell E A, \PRL {\bf 98}, 063201 (2007)}

\bibitem{ketterle_casimir}{Pasquini T A, Shin Y, Sanner C, Saba M, Schirotzek A, Pritchard D E and Ketterle W, \PRL {\bf 93}, 223201 (2004)}

\bibitem{vuletic_casimir}{Lin Y J, Teper I, Chin C and Vuleti\'c V, \PRL {\bf 92}, 050404 (2004)}

\bibitem{hinds}{Baumg\"artner F, Sewell R J, Eriksson S, Llorente-Garc\'ia I, Dingjan J, Cotter J P and Hinds E A, {\it Preprint} arXiv:1008.1252v1}

\bibitem{fattori}{Fattori M, D'Errico C, Roati G, Zaccanti M, Jona-Lasinio M, Modugno M, Inguscio M and Modugno G, \PRL {\bf 100}, 080405 (2008)}

\bibitem{kasevich}{Anderson B P and Kasevich M A, Science {\bf 282}, 1686 (1998)}







\bibitem{esteve} {Est\`eve J, Gross C, Weller A, Giovanazzi S and Oberthaler M K, {\it Nature} {\bf 455}, 1216 (2008)}



\bibitem{GrossNature2010} Gross C, Zibold T, Nicklas E, Est\'eve J and Oberthaler M K, {\it Nature} {\bf 464}, 1165 (2010).

\bibitem{treutlein} {Riedel M F, Böhi P, Li Y, Hänsch T W, Sinatra A and Treutlein P, {\it Nature} {\bf 464}, 1170 (2010)}

\bibitem{MaussangArxiv2010}{Maussang K, Marti G E, Schneider T, Treutlein P, Li Y, Sinatra A, Long R, Estève J and Reichel J, {\it Preprint} arXiv:1005.1922v1}

\bibitem{giovanetti} Giovanetti V, Lloyd S and Maccone L, {\it Science} {\bf 306}, 1330 (2004)

\bibitem{pezze}{Pezz\'e L and Smerzi A, \PRL {\bf 102}, 100401 (2009)}

\bibitem{shin}{Shin Y, Saba M, Pasquini T A, Ketterle W, Pritchard D E and Leanhardt A E, \PRL {\bf 92}, 050405 (2004)}

\bibitem{schumm} {Schumm T, Hofferberth S, Anderson L M, Wildermuth S, Groth S, Bar-Joseph I, Schmiedmayer J
  and Kr\"uger P, {\it Nature Physics} {\bf 1}, 57 (2005)}

\bibitem{AlbiezPRL2005} Albiez M, Gati R, Fölling J, Hunsmann S, Cristiani M and Oberthaler M K, \PRL {\bf 95}, 010402 (2005).

\bibitem{LevyNature2007} Levy S, Lahoud E, Shomroni I and J.~Steinhauer, {\it Nature} {\bf 449}, 579 (2007).

\bibitem{toronto}{LeBlanc L J, Bardon A B, McKeever J, Extavour M H T, Jervis D, Thywissen J H, Piazza F and Smerzi A, {\it Preprint} arXiv:1006.3550v2}

\bibitem{pezze2005}{Pezz\'e L, Collins L A, Smerzi A, Berman G P and Bishop A R, \PR A {\bf 72}, 043612 (2005)}

\bibitem{lee}{Lee C, \PRL {\bf 97}, 150402 (2006)}

\bibitem{huang}{Huang Y P and Moore M G, \PRL {\bf 100}, 250406 (2008)}

\bibitem{grond1}{Grond J, Hohenester U, Mazets I and Schmiedmayer J, \NJP {\bf 12}, 065036 (2010)}

\bibitem{braun}{Braunstein S L and Caves C M, \PRL {\bf 72}, 3439 (1994)}

\bibitem{chwed} Chwede\'nczuk J, Piazza F and Smerzi A, \PR A {\bf 82}, 051601(R) (2010)

\bibitem{laloe}{Mullin W J and Lalo\"e F, \PR A {\bf 82}, 013618 (2010)}

\bibitem{crlb1}{Helstrom C W, {\it Quantum Detection and Estimation Theory} (Academic Press, New York, 1976), Chap. VIII}
 
\bibitem{crlb2}{Holevo A S, {\it Probabilistic and Statistical Aspect of Quantum Theory} (North-Holland, Amsterdam, 1982)}

\bibitem{bach}{Bach R and Rz\c{a}\.zewski K, \PRL {\bf 92}, 200401 (2004)}

\bibitem{dobrzanski1}{Dorner U, Demkowicz-Dobrza\'nski R, Smith B J, Lundeen J S, Wasilewski W, Banaszek K, and Walmsley I A, \PRL {\bf 102}, 040403 (2009)}

\bibitem{dobrzanski2}{Kacprowicz M, Demkowicz-Dobrza\'nski R, Wasilewski W, Banaszek K and Walmsley I A, {\it Nature Photonics} {\bf 4}, 357 (2010) }

\bibitem{dobrzanski3}{Ko{\l}ody\'nski J and Demkowicz-Dobrza\'nski R, \PR A {\bf 82}, 053804 (2010)}

\bibitem{knysh}{Knysh S, Smelyanskiy V N and  Durkin G A, {\it Preprint} arxiv:1006.1645}

\end{thebibliography}
\end{document}